\documentclass[11pt]{article}
\usepackage{amssymb,amsmath}%,cite}
\usepackage[]{graphicx}
\usepackage{times}%,cite,w-thm}

\long\def\@makefntext#1{\parindent 0cm\noindent
\hbox to 1em{\hss$^{\@thefnmark}$}#1}

\begin{document}

\begin{titlepage}
\vspace{.5in}

%\begin{flushright}
%28 March 2008\\
%\end{flushright}
\vspace{.5in}
\begin{center}
{\Large\bf
 Gravitation equations,\\[.8ex] and space-time relativity}\\
\vspace{.4in}
{L.~V.~V{\sc erozub}\footnote{\it email: leonid.v.verozub@univer.kharkov.ua}\\
       %{\small\it Department of Physics}\\
       {\small\it Kharkov National University}}\\
       {\small\it Kharkov, 61103}\\{\small\it Ukraine}
\end{center}

\vspace{.5in}
\begin{center}
{\large\bf Abstract}
\end{center}
\begin{center}
\begin{minipage}{4.7in}
{\small
In contrast to electrodynamics,  Einstein's gravitation equations are not invariant with respect to a wide class of  the
mapping of field variables  which leave equations of motion of test particles  in a given coordinate system  invariant. 
It seems obvious enough that just these mappings should play a role of gauge transformations of the variables in  differential equations of gravitational field. 
We consider here in short a gauge-invariant bimetric generalisation of the Einstein  equations which does not contradict availabel observation data.   Physical interpretation of the bimetricity    based on relativity of space-time with respect to used reference frame,    following  conceptually  from old  Poincar\'{e}    fundamental ideas,  is proposed..
}
\end{minipage}
\end{center}
\end{titlepage}
\addtocounter{footnote}{-1}

The relativistic differential equations of motion of charges in  electromagnetic field are invariant with respect to some transformations of  the field four-potential. For this reason it is naturally that  Maxwell equations are also invariant with respect to these transformations. Similarly, the differential equations of motion of test particles in gravitational field in Einstein's theory in a given coordinate system
 are   invariant with respect to the following transformations of the Christoffel symbols   $\Gamma_{\beta\gamma}^{\alpha}$ 
\footnote{Greek indexes run from $0$ to $3$}  \cite{Weyl}-\cite{Eisenhart} of a Riemannian space-time $V$ :
\begin{equation}
\label{GammaGeodesTransformations}
\overline{\Gamma}_{\beta\gamma}^{\alpha
}(x)=\Gamma_{\beta\gamma}^{\alpha}(x)+\delta_{\beta}^{\alpha}\ \phi_{\gamma
}(x)+\delta_{\gamma}^{\alpha} \phi_{\beta}(x),
\end{equation}
where $\phi_{\beta}(x)$  are an arbitrary differentiable vector-function.
It is the most easier for seeing,  if the geodesic equations are written in the form
\begin{equation}
\ddot{x}^{\alpha}+(\Gamma_{\beta\gamma}^{\alpha}-c^{-1}\Gamma_{\beta\gamma
}^{0}\dot{x}^{\alpha})\dot{x}^{\beta}\dot{x}^{\gamma}=0.
\label{EqMotionOfTestPart}%
\end{equation}
where the dot denotes differentiation with respect to  $t=x^{0}/c$.  

The Ricci and metric tensors    also are not invariant  under  above self-mapping of  Riemannian space-time  
  which leave geodesic lines invariant. (They named geodesic mappings).

In contrast to Maxwell theory, Einstein's equations are not invariant under these transformations \cite{Petrov},
 although  it  seems reasonable to suppose that  just they are the transformations that have to play a role of gauge transformations of  field variables in  differential  equations of gravitation. It is a very strange fact, especially  taking into account that physical consequences  resulting from Einstein's equations  agree very closely  with all observations. data.  

The most natural explanation of such situation is that if there are more correct gravitation equations than the Einstein ones,  they may  differ markedly from the last equations only at very strong field, close the Schwarzschild radius, where we have not yet    firm evidences of  validity of physical consequences of the Einstein equations. \footnote{ It follows from (\ref{GammaGeodesTransformations}) that the components $\overline{\Gamma}^{i}_{00}=\Gamma^{i}_{00}$. Therefore, in Newtonian limit 
geodesic-invariance is not an essential fact. Therefore, now  we deal with a relativistic effect.}

  The simplest object  being geodesic invariant is the Thomas symbols
 \cite{Thomas}:
\begin{equation}
\Pi_{\alpha\beta}^{\gamma}=\Gamma_{\alpha\beta}^{\gamma}-(n+1)^{-1}\left[
\delta_{\alpha}^{\gamma}\ \Gamma_{\beta}+\delta_{\beta}^{\gamma}%
\ \Gamma_{\alpha}\right]  \; \label{ThomasSymbols}.
\end{equation}
where $\Gamma_{\alpha}=\Gamma_{\beta\alpha}^{\beta}$.

The simpest geodesic-invariant generalisation of the vacuum Einstein equations are 
\begin{equation}
{\mathcal{R}}_{\alpha\beta} =0,
\label{equations0}
\end{equation}
where  ${\mathcal{R}}_{\beta\gamma}$ is an object which is formed by the gauge-invariant
Thomas symbols the same way as the Ricci tensor is formed out of the Christoffel symbols.

However, the problem is that  $\Pi_{\alpha\beta}^{\gamma}$, as well as  ${\mathcal{R}}_{\beta\gamma}$, is not a tensor.

This problem can be solved,  if we will be consider all 
 geometrical objects in $V$ as some objects in the  Minkowski  space-time by analogy with   Rosen's 
bimetric theory \cite{Rosen}.
 It means that we must replace all derivatives in geometrical objects o the f Riemannian space-time by the covariant ones defined in the Minkowski space-time. After that,  in an arbitrary coordinate system 
we obtain instead  $\Gamma_{\beta\gamma}^{\alpha}$ a tensor object 
$D_{\beta\gamma
}^{\alpha}=\Gamma_{\beta\gamma}^{\alpha}-\overset{\circ}{\Gamma}_{\beta\gamma
}^{\alpha}$
,  where $\overset{\circ}{\Gamma}_{\beta\gamma
}^{\alpha} $ is  Christoffel symbols of Minkowski space-time   $E$ in  used coordinate system.
In like manner we obtain instead  Thomas symbols a geodesic-invariant (i.e. gauge-invariant) tensor 
\begin{equation}
B_{\beta\gamma}^{\alpha}=\Pi_{\beta\gamma}^{\alpha}-\overset{\circ}{\Pi
}_{\beta\gamma}^{\alpha},
\end{equation}
where $\overset{\circ}{\Pi
}_{\beta\gamma}^{\alpha} $ are  the Thomas symbols in the Minkowski space-time.
This tensor   must play a  role of  a strength tensor of gravitational field.
Now,  using the identity $B_{\alpha\beta}^{\beta}=0 $,     we obtain  instead  ( \ref{equations0})
a geodesic-invariant bimetric equation which can be written in the form
\begin{equation}
\nabla_{\alpha}B_{\beta\gamma}^{\alpha}-B_{\beta\delta}^{\epsilon}%
B_{\epsilon\gamma}^{\delta}=0, \label{MyVacuumEqs}%
\end{equation}
where $\nabla_{\alpha}$ denotes a covariant derivative in $E$.
Some generalisation of the Einstein's equations can be  obtained and for  the case of matter presence.

Evidently, these bimetric equations may be true if   both the space-times,
$V$ and $E$, have some physical meaning. But how these two physical space-time can coexist?
%Reflections about this problem lead us to one of the deepest problems of space-time theory.

An attempt to answer this question  leads us to  discussion of a fundamental   problem of  relativity of space-time with respect to  properties of used measuring instruments. A  fresh look at  Poincar\'{e}  old  well known  results allows 
 to obtain  conclusions which  revise our understanding of  geometrical properties of space-time .

At beginning of  20th century 
Poincar\'{e}      showed  \cite{Poincare}      that
only an aggregate " geometry + measuring instruments" has a physical 
 meaning,  verifiable by experiment, and  it makes no sense to assert that one or other geometry of physical space 
in itself is true. In fact,  the equations of Einstein, 
is  the  first attempt to fulfil    ideas of Berkeley - of Leibnitz - Mach about  space - time relativity. Einstein's equations clearly show that there is a relationship between properties of space - time and matter  distribution . However Poincare's ideas testify that  space and time relativity is not restricted only to dependency  of space-time geometry on matter distribution. The space-time geometry  also depends  on properties of measuring instruments. However, a choice of certain 
properties of the measuring instruments is  nothing more than the choice of certain
frame of reference, which just  is  such a physical device by means of which we
test properties of space-time.  Consequently, one can  expect that there is a relationship between the metric of 
space- time and  a used reference frame.

 A step towards the implementation of such idea is considered  in \cite{Verozub08a}.
By a non-inertial frame of reference (NIFR) we mean the frame, the body of reference
of which is formed by point masses moving in an inertial frame of reference (IFR)  under the effect of a
 force field. 
By proper frame of reference of a given force field we mean the NIFR, the reference body of which is formed by  
point masses moving under the effect of the force field. We postulate that  space-time in IFRs
is the Minkowski one,   in accordance with special relativity.  Then,  above definition of  NIFRs allows to find line element of space-time in PFRs. 

 Let   $\mathcal{ L}(x,\dot{x})$ be  Lagrangian describing in an IFR  the motion of  point particles with  masses  $m$ 
forming  the reference body of  a NIFR .
In this case can be sufficiently clearly argued  \cite{Verozub08a} 
that the line element  $ds$  of space-time   is given by
\begin{equation}
ds=-(mc)^{-1} \, dS(x,dx),
\label{dsMain}%
\end{equation}
where  $S=\int{ \mathcal{ L}(x,\dot{x}) dt}$ is the action describing  the motion of  particles of the reference body  in the Minkowski space-time.
Therefore, properties of space-time in PFRs are entirely determined by
properties of used frames in accordance with the  Berkeley-Leibnitz-Mach-Poincar\'{e}
 ideas of relativity of
space and time. 

We can illustrate the above result by some examples.

1. The reference body consists of noninteracting electric charges in a
constant homogeneous electromagnetic field.   The
Lagrangian describing the motion of  charges with masses $m$ is of the form:
\begin{equation}
L=-mc^{2}(1-v^{2}/c^{2})^{1/2} - \phi_{\alpha}(x)  dx^{\alpha}),
\label{LagrangGarge}.
\end{equation}
where  $\phi$ is a vector function,  $c$ is the speed of light, and $v$ is the spatial velocity. 
Then, according to (\ref{dsMain}),   the line element of space-time in the PFR is given by
\begin{equation}
ds= d\sigma+f_{\alpha}(x)dx^{\alpha} \label{dsRanders}%
\end{equation}
where  $f_{\alpha}=\phi/m$ is a vector field, and $d\sigma $ is the line element of the Minkowski space-time. Consequently,  space-time in PFRs of electromagnetic field  is Finslerian. In principle, we can use both   traditional and geometrical description,  although  the last in this case is rather too complicate.  

2. Motion of  an ideal isentropic fluid can be considered as the motion of macroscopic small elements (``particles'')
of an arbitrary  mass $m$,    which is described   by  the Lagrangian 
 \cite{Verozub08b}%
\begin{equation}
L=-mc \left(  G_{\alpha\beta} \dot{x}^{\alpha} \dot{x}^{\beta}\right)^{1/2}
\label{Lagrangian_in_V},%
\end{equation}
where 
$w$  is   enthalpy per unit volume, 
$G_{\alpha\beta}=\varkappa^{2}\eta_{\alpha\beta}$ , $\varkappa=w/\rho c^{2}$, $\rho=m n$, 
$m$ is the mass of the particles,
$n$ is  the
particles number density,  and
$\eta_{\alpha\beta}$ is the metric tensor in the Minkowski space-time. According
to  (\ref{dsMain}) the line element of space-time  in the NIFR  is
given by
\begin{equation}
ds^{2}=G_{\alpha\beta}dx^{\alpha}dx^{\beta} . \label{ds2}%
\end{equation}
Therefore, the motion of the particles can be considered as occurring under the effect of a force field. (In non-relativistic case it is a pressure  gradient).
Space-time in the PFR of this force field is Riemannian,  and  conformal to Minkowski space-time. The motion of the above particles does not depend on theirs masses. We can use both traditional and geometrical description.  In some cases such geometrical description  is preferable.

3. Suppose   that  in the Minkowski space-time  the Lagrangian describing  the motion of test particles  of mass $m$ in  a tensor field  $g_{\alpha\beta}$ is of the form
\begin{equation}
L=-mc[g_{\alpha\beta} \;\dot{x}^{\alpha}\;\dot{x}^{\beta}]^{1/2},
\label{LagrangianThirr}
\end{equation}
where $\dot{x}^{\alpha}=dx^{\alpha}/dt$.
According to (\ref{dsMain} ),  the line element of space-time in the PFR is given by
\begin{equation}
ds^{2}=g_{\alpha\beta}\;dx^{\alpha}\;dx^{\beta}.
\end{equation}
Space-time in PFRs of this field is Riemannian, and  motion of test particles do not depend of their masses.
It is natural to assume that in this case we deal with a gravitational field. 

The bimetricity in this case has a simple physical meaning.  
Disregarding the
rotation of the Earth, a reference frame, rigidly connected with the Earth
surface, can be considered as an IFR. An observer, located in
this frame, can describe the motion of freely falling identical point masses as
taking place in Minkowski space-time under the effect of a  force
field. However,  for another observer  which is  located in the PFR 
 the reference body of which is formed by these freely falling
particles, the situation is another.  
 Let us assume that the observer is deprived
of the possibility of seeing the Earth and stars. 
Then, from his point of view,  the the point masses formed the reference body  of the PFR are points of his physical space, and all events occur   in his space-time. Consequently,   accelerations of these point masses must be equal to zero  both in nonrelativistic and relativistic meaning  .
 However, instead of this, he observes a change in distances between
these point masses in time.  Evidently, the only
reasonable explanation for him is the interpretation of this observed
phenomenon as a manifestation of the deviation of geodesic lines in some
Riemannian space-time of a nonzero curvature. Thus, if the first observer,
located in the IFR, can postulate that space-time is flat, the second
observer, located in a PFR of the force field, who proceeds from relativity of
space and time, already in the Newtonian approximation
\textit{is forced} to consider space-time as Riemannian with curvature other
than zero.

To obtain physical consequences from (\ref{MyVacuumEqs})  it is convenient to select the gauge condition
\begin{equation}
 Q_{\alpha}=\Gamma_{\alpha\beta}^{\beta}-\overset{\circ}{\Gamma^{\beta}}_{\alpha\beta}=0.
\label{AdditionalConditions}
\end{equation}
 At such gauge condition (which does not depend on coordinate system) eqs. (\ref{MyVacuumEqs}) conside with the vacuum Einstein's equations . 
Therefore, for solving many  problems it is sufficiently to find solution of  thvacuumum Einstein equations in the Minkowski space-time   (in which $g_{\alpha\beta}(x)$ is simply a tensor field) at the  condition $Q_{\alpha}=0 $ .

From the point of view of the observer located in an  IFR and 
studying the  gravitational field of a remote compact object of mass $M$, the space - time  is flat.  The spherically-symmetric   solution of the equations (\ref{MyVacuumEqs} ) for the  point central object very little
 differs   from the solution   in  general relativity, if the distance from the center $r$ is much more that thSchwarzschildld radius  $r_{g}$. However these solutions  in essence differ as   $r$  is of the order of $r_{g}$ or less than that.    The solution  in flat space   has no singularity at centre and the event horizon at $r=r_{g}$.  
\begin{figure}[bhp]
\includegraphics[width=8cm,height=6cm]{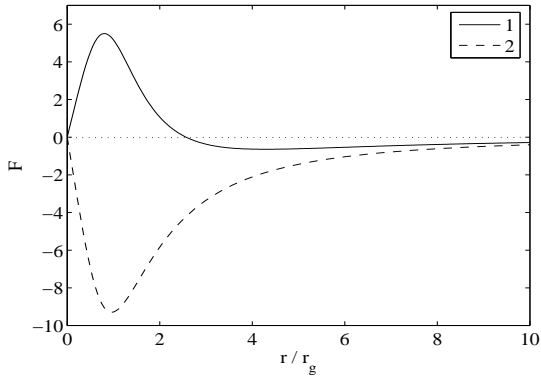}
\caption{The gravitational force (arbitrary
units) affecting freely  particles (curve 1) and rest-particles (curve
2) near an attractive point mass.}%
\label{Force}
\end{figure}
Fig. 1 shows the plots  of the gravitational force  $F=m \ddot{x}^{\alpha}$
acting  on rest-particle  of mass $m$  and 
on a freely falling test particle     as a functions of  $r$.  It follows from the figure , that in the first case $F$ tends to zero when $r\rightarrow 0$.   In the second case, as particle approach to the Schwarzschild  radius,  the force changes the sign and becomes repulsive. 

These unexpected 
peculiarities of the gravitational force can be tested by observations.  The peculiarity of of the static force leads to the  possibility of the existence of supermassive compact  objects without  event horizon. Such objects can be identified with supermassive  compact  objects at centres of galaxies.  \cite{Verozub06a}.

The unusual properties of the force acting  on freely moving  particles  near  the Schwarzschild radius  give rise to some  observable effects in cosmology  because it is well-known  that the radius of an observable part of the Universe  is of the order of the Schwarzschild radius  of all observed mass. 
It yields a natural explanation of a deceleration of the Universe expansion \cite{Verozub08a}.


\begin{thebibliography}{ }

\bibitem{Weyl}  H.\,  Weyl,  G\"{o}ttinger Nachr., 90 (1921).

 \bibitem {Thomas} T.\,   Thomas, The differential invariants of generalized spaces, 
(Cambridge, Univ. Press) (1934).

\bibitem {Eisenhart} L.\,  Eisenhart, Riemannian geometry,  (Princeton,  Univ. Press) (1950).

\bibitem {Petrov}A.\,  Petrov, Einstein Spaces , (New-York-London, Pergamon Press. (1969).

\bibitem {Poincare} H.\,  Poincar\'{e}, Derni\`{e}res pens\'{e}es, (Paris, Flammarion) (1913)

\bibitem{Verozub08a} L.\. Verozub, Ann. Phys. (Berlin), \textbf{17}, 28 (2008)

\bibitem{Verozub08b} L.\, Verozub, Int. J. Mod. Phys. D, \textbf{17}, 337 (2008)

\bibitem {Rosen} N.\,  Rosen, Gen. Relat. Grav., \textbf{4}, 435 (1973).

\bibitem{Verozub06a} L.\, Verozub, Astr. Nachr., \textbf{327}, 355 (2006)

\end{thebibliography}
\end{document}